\documentstyle[12pt]{article}
\newcommand{\be}{\begin{equation}}
\newcommand{\ee}{\end{equation}}
\begin{document}
YITP-00-41
\hspace{10cm}
\today
\\
\vspace{3cm}
\begin{center}
{\LARGE The Supersymmetric Kink via Higher-Derivative and Momentum Cut-Off
Regularization Schemes}
\\  \vspace{2cm}
{Andrei Litvintsev \footnote{e-mail: litvint@insti.physics.sunysb.edu} \\
 { \it C.N.Yang Institute for Theoretical Physics, SUNY at Stony Brook}, \\
{\it Stony Brook, NY 11794 } }
\abstract{\it{
 We study the supersymmetric kink with higher derivative and
momentum cut-off regularization schemes.
We introduce the new momentum cut-off regularization scheme which we call ``generalized momentum cut-off''.  
A new, explicit 
computation for the central charge anomaly for this 
scheme is described in detail.
The calculation of the first order mass corrections for the bosonic and 
supersymmetric kink
within the momentum cut-off is presented in \cite{we}, so that one can
compare the one-loop  central charge and mass computed 
independently within the same
regularization setup. We confirm that the BPS bound is saturated in one 
loop level. Thus the Wilsonian momentum cut-off
regularization scheme is rehabilitated as a bona fide procedure for
computing quantum corrections in the topologically nontrivial backgrounds.
We lead the reader to the idea that a consistent regularization 
in general does 
not only assume that one
regulates loops properly, but also requires caution in defining 
the total number of modes involved in the quantization of the theory.
We also study the higher-derivative regularization
scheme of \cite{shifman} in great detail. We extend the Noether method 
for the case when the  
higher-derivatives are present in the Lagrangian. The extensive discussion of the technical aspects and the consistency of the
higher-derivative scheme is given. 
We show that higher-derivative regularization gives, in general, a nonlocal 
topological current, which leads to the correct value of the anomaly in the 
central charge. We also comment on the status of the dimensional regularization approach to the computation of the central charge anomaly and kink mass.
} } 
\end{center}
\newpage

\section{Introduction}

The problem of the quantum corrections to the mass of the
(1+1) dimensional supersymmetric kink - the topologically 
nontrivial configuration in a system with the Lagrangian
\begin{equation}
\label{boldlagrangian}
L=\frac{1}{2} \left\{ \partial_\mu \phi   \partial^\mu
\phi + \bar{\psi}   i
  {\not \partial} \psi + F^2 + 2 W^\prime(\phi) F - W^{\prime \prime} (\phi) 
\bar{\psi} \psi \right\} ,
\ee
with 
\be
\label{potencial}
W^\prime(\phi) =  \frac{m^2}{4 \lambda} - \lambda \phi^2 
\ee
 has had more than twenty years 
of history. Different perturbative computations gave different answers and
only recently most of the loose ends of different approaches were 
tide up to form
the consistent picture. 
A review of this picture was given in \cite{we}. 

The mass of the bosonic or supersymmetric kink is defined as being the difference between the 
vacuum expectation values (VEVs) of Hamiltonian in the nontrivial and trivial backgrounds. 
Both of these VEVs are divergent and the difference is meaningless unless one uses some regularization
scheme to make these sums finite (or at least logarithmically divergent).  
A general expression for the mass  is clear
from the following scaling property: under the rescaling 
$$
\breve{z}=\Lambda^{-1/2}z, 
\breve{t}=\Lambda^{-1/2}t,
\breve{\phi}=\Lambda^{1/2}\phi,
\breve{\psi}=\Lambda^{3/4}\psi,
\breve{\hbar}=\Lambda \hbar,
\breve{m}=\Lambda^{1/2}m
$$
the VEV of Hamiltonian scales as 
$$
\langle H(m,\lambda) \rangle_\hbar = \frac{1}{\Lambda^{3/2}} \langle H( \Lambda^{1/2}
m, \lambda) \rangle_{\Lambda \hbar} \  ,
$$
which implies (we also use that both $\lambda$ and $m$ in (\ref{potencial}) both have the dimensions of mass) the following form of this VEV
$$\langle H(m,\lambda ) \rangle_\hbar = a \frac{m^3}{\lambda^2}+ b \hbar m +
c \hbar^2 \frac{\lambda^2}{m}+O(\hbar^3) $$
where $a$,$b$,$c$ are some numerical coefficients. 
For the last twenty five years, there were a lot of controversal computations of the value of $b$ in the literature
(see \cite{peter} for further references).  
The agreement on this value was reached only recently after the introduction
of the derivative regularization scheme in ref. \cite{misha}. The generally accepted values
of $b$ for bosonic and susy kinks are 
\be
b(bos) =  -m  \left( \frac{3}{2 \pi} -
\frac{\sqrt{3}}{12} \right); \hspace{1cm}
b(susy) = Z^{(1)}(susy) = - \frac{ m}{2 \pi} 
\label{ura1}
\ee

Once the mass value was settled,
the next important question, the BPS bound saturation, was addressed by \cite{misha}.
Though mentioning the word ``anomaly'', \cite{misha} failed to check the saturation.
The real understanding came with \cite{shifman}, where it was realized that there is
the quantum ultraviolet anomaly in the central charge which is responsible
for the exact saturation of the BPS bound. In our opinion, the pioniering 
work \cite{shifman} had some technical mistreatments in the implementation of the higher
derivative and dimensional regularization schemes.
It is the goal of 
the present work to clarify the issue of the anomaly in the central charge
by, at one hand, giving a new computation of the anomaly in the new
generalized momentum cut-off scheme, and, on the other hand, by presenting
a different, self-consistent computation of an anomaly with the higher-derivative scheme of ref. \cite{shifman}. 

We start with the momentum cut-off regularization scheme.
This is one of the most popular schemes for the perturbative calculations.
Nevertheless, previous, naive implementations of this scheme did not
disclose the anomaly in the central charge. We solve this problem. We
notice that
it is important to regularize the 
problem consistently {\it before} we start to do any computations. This is
a general principle that must be used in any regularization procedure: 
regularization scheme is a part of the model and should be put in it from the 
very start. For the case of momentum cut-off scheme, this rule says that
one should declare that only plane waves with momenta up to 
cut-off are considered
in the theory, put into path integrals and used to expand quantum fields
in second quantized form. In particular, the
propagators and the Dirac delta functions 
in the canonical commutation relations
should contain only the modes with momenta smaller 
than a cut-off, that is we must impose a
cut-off on Dirac delta functions as well as on loops. 
Of course, this cut-off should
be exactly the same as in the propagators. Nevertheless, we first
put two different cut-offs, one is for quantum fields, another is for
delta functions. ( This will allow us to understand easily some results of the 
higher-derivative regularization approach). We observe that there are only
three different results for the anomaly that are possible, depending
on which cut-off is larger, or both are equal. For equal cut-offs
we reproduce the correct value of the anomaly. 

We would like also to point out that in paper \cite{we} we develop an
accurate momentum cut-off treatment of the Casimir energy sum, thus
computing directly the mass of the supersymmetric kink. 
To the best of our knowledge
this is the first time when the anomaly in the central charge and the quantum
mass are computed independently within the framework of the 
same regularization setup.
In our case, this setup is the momentum cut-off scheme
\footnote{ We are aware of paper \cite{vse} which claims to do the same. We will describe later in this paper why
one should understand \cite{vse} as only being valid in the context of the dimensional regularization. Also,
that paper does not refer to the central charge anomaly, and our understanding is that it just reduces the 
computation of central charge VEV to the computation of the hamiltonian VEV plus zero, arising from the VEV
of the susy generator component squared.}.

Another scheme, the higher-derivatives regularization, was proposed by \cite{shifman}. 
All the inputs of this scheme are written in the superspace Lagrangian, 
thus the
supersymmetry is manifestly preserved by this regularization. 
We study this regularization scheme in great detail.
We develop the Noether method for higher-derivative Lagrangian. 
The resulting supersymmetry current is not unique. 
Using Noether method with susy 
tranformations one can compute a whole family of currents $j^\mu$, 
which differ from
each other by a trivially conserved term of the form 
$\epsilon^{\mu \nu} \partial_\nu A(x,t)$. 
Notice that, in general, not only
fermionic anticommutators contribute to the central charge, but also bosonic 
canonical commutation relations give a finite first-order contribution. 
These two terms always add to give a proper value for the anomaly. The fact 
that the above mentioned nonuniqueness could not possibly change the central charge value also follows from
the general form of the susy algebra, as it was pointed to us by \cite{pismo}.
Nevertheless, we claim that only one of the currents presented in this paper
can be used for the anomaly computation. The reason is that the other currents 
are not properly regulated by the theory, vacuum averages of this currents
are divergent, though the space integration leaves only finite contributions
to the anomaly. 
One must pick some very special current
which we call $j_{OK}^\mu$ to have the loops regulated properly.
This current is different from the one used in \cite{shifman}.   
We believe that we present an accurate
and consistent treatment of the model and we compute
the correct anomaly value from the properly regulated current.

The correct result for the anomaly turns out to be exactly equal
to the one loop mass correction, computed by \cite{misha}, and then checked by
different methods in \cite{vse} and \cite{dune}. That is, the BPS bound is saturated
in one loop, which confirms the claim made in \cite{we,fred} 
(after reanalysing arguments of \cite{shifman}) that there is multiplet
shortening in $1+1$ dimensions and the BPS bound is saturated in 
any number of loops.   

\section{Supersymmetric kink in momentum cut-off regularization}
We begin with some definitions\footnote{
Note that our notations are the same as in ref. \cite{shifman} and they differ from the conventions of ref. \cite{we}.
}. Our metric is $(+-)$, $\gamma^0=\sigma_2$,
$\gamma^1=i \sigma_1$, and $\bar{\psi}=\psi^T \gamma^0$. The action for a scalar 
superfield $\Phi$ is 
\begin{equation}
\label{saction}
S=i \int d^2\theta d^2x \left\{ \frac{1}{4} \bar{D}^\alpha \Phi
 D_\alpha \Phi + W(\Phi) \right\}
\end{equation}
where we will take 
$
\label{sup}
W(\Phi)=\frac{m^2}{4 \lambda} \Phi - \frac{\lambda}{3} \Phi^3
$. 
The corresponding Lagrangian in component fields is given by 
(\ref{boldlagrangian}). 
The action is invariant under the supersymmetry transformations given by
$$
\delta \phi = \bar{\epsilon} \psi
$$
\be 
\label{susy}
\delta \psi = - i (\not \partial \phi) \epsilon + F \epsilon
\ee $$
\delta F = - i \bar{\epsilon} (\not \partial \psi)
$$
The nontrivial classical solution of the corresponding equations of 
motion is the kink 
\be
\label{boldkink}
\phi_0(x)=\frac{m}{2 \lambda} \tanh \frac{mx}{2}
\ee

To study the quantum theory with the action (\ref{saction}) one needs to regulate and renormalize. 
Only mass renormalization is required for this model. We will adopt the most popular renormalization
scheme, which requires that the one-loop tadpole graphs are exactly cancelled by the mass counterterm, 
see ref.\cite{we} for details. 
In this chapter we pick the momentum cut-off regularization scheme
to do that, i.e. in the quantum fields we are going to keep 
only modes with the momenta 
smaller than the cut-off scale $\Lambda$. Note that, in general, this implies  
that the classical solution (\ref{boldkink}) is
no longer correct - one needs to remove from it the modes with the momenta higher than cut-off $\Lambda$. 
Nevertheless, this won't influence further discussion because such a 
correction would vanish when
the cut-off is pushed to infinity. (This will be clear from the following discussion and eq. (\ref{metka1})
if one keeps in mind that only the antisymmetric modes $\sin kx$ contribute to the Fourier 
expansion of (\ref{boldkink}) ). We impose the
equal time commutation relations 
\be \left[ \phi(t,x),\partial_0 \phi(t,x^\prime) \right] =
i \hat{\delta}(x-x^\prime) 
\hspace{1.8cm}  
\left\{\psi_\alpha(t,x),\psi^{T \beta}(t,x^\prime)\right\}=\delta_\alpha^\beta
\hat{\delta}(x-x^\prime)
\ee
where we keep a finite number of Fourier modes, namely the modes with momenta below cut-off $K$ in delta function, i.e.
\be
\hat{\delta}(x) = \int_{-K}^K \frac{dq}{2\pi} \exp(iqx)
\ee
The theory is consistent only if $\Lambda=K$. Nevertheless, we wish to keep
these two parameters different. The result in terms of both of these parameters
will describe the nature of problems that one may encounter in much more 
complicated schemes.

Using Noether method, one can compute the supersymmetry current
\begin{equation}
j^\mu = ( \partial_\nu \phi) \gamma^\nu \gamma^\mu  \psi + i W^\prime 
\gamma^\mu \psi
\end{equation}
The supersymmetry charge  $Q$ is defined as the space integral of the time
component of this current. We are computing the supercharge algebra
\begin{equation}
\label{alg}
\{ Q_\alpha, \bar{Q^\beta} \} = 2 (\gamma_\nu)_\alpha^\beta P^{\nu}+2 i (\gamma^5)_\alpha^\beta Z
\end{equation}
where we have introduced the central charge $Z$ and the momentum $P^\nu$. We are going to compute
the VEV of $Z$ in one loop. Consider the anticommutator in (\ref{alg}).
From counting the number of $\gamma$-matrices it is clear that 
 the central charge 
contributions come from two sources: \\
1) from the fermionic anticommutator 
of fermionic field of the first term in the current with the fermionic field
of the second term in the current, and then computing bosonic loop; \\
2) from the bosonic commutator of $\partial_0 \phi$ term in the first term of the
current with the bosonic field in the second term of the current.

For the case 1) one computes
\be 
Z_{bos}=\int_{-L}^{L} dx \int_{-L}^{L} dx^\prime W^\prime(x)
 \partial_{x^\prime} \phi(x^\prime) \hat{\delta}(x-x^\prime)
\ee
This is an operator expression. (We suppress the time dependence of fields
for the compactness of formulas). We now write the field as
$\phi(x)=\phi_0(x)+\eta(x)$  and compute the VEV of this operator
in one loop. This gives 
 $$ \langle Z_{bos} \rangle =Z_{0}+\int_{-L}^{L} dx \int_{-L}^{L} dx^\prime
 W^{\prime \prime \prime}(x)
 \partial_{x^\prime} \phi_0(x^\prime) \hat{\delta}(x-x^\prime) \langle
\eta^2(x) \rangle + 
$$ \be \int_{-L}^{L} dx \int_{-L}^{L} dx^\prime 
W^{\prime \prime}(x)
\langle \eta(x) \partial_{x^\prime} \eta(x^\prime) \rangle \hat{\delta}(x-x^\prime)
\label{metka1}
\ee
where $Z_0$ is the classical term, and the rest is the one loop correction.
The second term in this expression can be written as a total derivative
of $W^{\prime \prime}$, and we recognize the regular logarithmic term
which comes from the second order Taylor expansion when one computes the 
one loop correction to $Z=W$ term. This term can be removed by the 
renormalization. 

The third term in (\ref{metka1}) is responsible for the half of the
total anomaly (another half being given by the case 2) ). 
The correlator in this term is an odd function of $(z-z^\prime)$.
So, we need to compute the integral
\be
\label{int1}
\int_{-L}^L dx \int_{-L}^{L} dx^\prime W^{\prime \prime}(x) i\int_{-\Lambda}^{\Lambda}
\frac{d^2p}{(2\pi)^2} \frac{i p_x}{p^2} \exp(ip_x(x-x^\prime)) \int_{-K}^{K}
\frac{dq}{2\pi} \exp(iq(x-x^\prime))
\ee
We find for this integral in the limit $L \to \infty$ 
\be
\label{polotveta}
\frac{W^{\prime \prime}(\infty)}{4 \pi} \left( 1+{\rm sign} (\Lambda-K ) \right)
\ee
Of course, it is easy to compute this integral by computer. We feel that 
the reader, who likes to have all the computations under the control, 
may have some technical problems computing this integral and 
that is why we devote the appendix A to this computation.

Now, let us move to the case 2). Consider what one would get from commutators
of $\partial_0 \phi$ with $\phi$ and $\partial_x \phi$ in the supersymmetry charge.  
The $\gamma^5$ part results in the following integral for 
the central charge contribution
$$ \int_{-L}^{L} dx \int_{-L}^{L} dx^\prime  \left\{ i \partial_{x^\prime} 
\hat{\delta}(x-x^\prime) \langle \bar{\psi}(x^\prime) \psi(x) \rangle 
+ W^{\prime \prime}(x) \hat{\delta}(x-x^\prime) \langle \bar{\psi}(x^\prime)
\gamma^1 \psi(x) \rangle \right\}
$$ 
The first term here is zero (the $p$ term in the fermionic propagator is 
killed by trace of gamma matrix, the $m$ term is just zero due to the
parity properties). The $\gamma^1$ in the second term picks the
$x$ component of the loop momentum and we get exactly the same mathematical
expression for the second term as one in the (\ref{int1}). 

This way we observe that the total result for the central charge
anomaly is 
\be
\label{ress1}
\frac{W^{\prime \prime}(\infty)}{2 \pi} \left( 1+{\rm sign} (\Lambda-K ) \right)
\ee
and bosonic and fermionic loops contribute into it equally.

The formula (\ref{ress1}) has interesting properties.
What if we used the full delta function instead of $\hat{\delta}$?
Then it would remove one of the coordinate integrals from the very start, 
and we would loose the anomaly! This is generally true for any $K>\Lambda$,
the anomaly computed in such a scheme is zero. This is the reason why
this anomaly has been overlooked by earlier papers. 
On the other hand, if we would under-regulate propagators, i.e. put
$\Lambda>K$, then the anomaly which is  computed in this way would be twice 
the one, that we present here. It is needless to say that both
cases result in the internally inconsistent theory.  

Thus we conclude that for properly regularized theory,
i.e. for $\Lambda=K$, the central charge anomaly is $$\frac{W^{\prime \prime}(\infty)}{2 \pi}$$ and the BPS bound is saturated.

The scheme discussed here violates supersymmetry. It would be useful
to repeat this analysis for the manifestly supesymmetric regularization.
We are doing it in the next chapter for the regularization scheme, proposed by [2].

\section{Supersymmetric kink in higher-derivative regularization}

Consider the regularized  action for a scalar 
superfield $\Phi$ in the superspace 
\begin{equation}
S=i \int d^2\theta d^2x \left\{ \frac{1}{4} \bar{D}^\alpha \Phi
 \left( 1 - \frac{\partial^2_z}{M^2} \right) D_\alpha \Phi + W(\Phi) \right\}
\end{equation}
where $M$ is a regulator mass. This scheme does not have a sharp cut-off
scale $\Lambda$, as the model before, but it introduces
the characteristic scale $M$ at which the momentum loops are becoming suppressed.
This is a smooth cut-off counterpart of the sharp momentum cut-off scheme.
The corresponding Lagrangian in component fields is
\begin{equation}
L=\frac{1}{2} \left\{ \partial_\mu \phi  \left( 1 - \frac{\partial^2_x}{M^2} \right) \partial^\mu
\phi + \bar{\psi}  \left( 1 - \frac{\partial^2_x}{M^2} \right) i
  {\not \partial} \psi 
\right.
\end{equation}
$$+ F  \left.
 \left( 1 - \frac{\partial^2_x}{M^2} \right) F + 2 W^\prime(\phi) F - W^{\prime \prime} (\phi) 
\bar{\psi} \psi \right\} 
$$
The action is still invariant under the supersymmetry transformations 
(\ref{susy}). 
The field equations are satisfied automatically by $\psi_\alpha=0$ and time independent $\phi(x)$ which
satisfy
\begin{equation}
\label{kink}
 \left( 1 - \frac{\partial^2_x}{M^2} \right) \partial_x \phi \mp W^\prime (\phi) =0
\end{equation}
One can observe directly that this equation is really needed to minimize potential energy of the
system. Just multiply left hand side of (\ref{kink}) by
$$
\partial_x \phi \mp \left( 1 - \frac{\partial^2_x}{M^2} \right)^{-1}  W^\prime (\phi)
$$
and the result is equal to a potential energy term of Lagrangian up to total derivatives.

The nontrivial solution of (\ref{kink}) with $-$ sign we call supersymmetric kink with higher-derivative regularization. 

Now we compute the supersymmetric kink shape up to the first order in $1/M^2$. The field equation
$$
\left( 1-\frac{\partial^2_x}{M^2} \right) \partial^2_x \phi - W^{\prime \prime} \left( \frac{1}{1-\frac{\partial^2_x}{M^2}} W^\prime \right) = 0
$$
has the nontrivial solution of the form
$$
\phi(x) = \frac{m}{2 \lambda} \tanh \frac{m x}{2} + \frac{1}{M^2} f(x)+O(\frac{1}{M^4})$$
after change of variables $\zeta=\tanh \frac{m x }{2}$ we arrive at
the following equation for $f(\zeta)$
$$
\frac{\partial}{\partial \zeta} (1-\zeta^2) \frac{\partial}{\partial \zeta} f(\zeta) +6 f(\zeta) - \frac{4}{1-\zeta^2} f(\zeta)+ \frac{m^3}{2 \lambda} (9 \zeta^3-5 \zeta)=0
$$
which has the exact solution of the form
$$
f(\zeta)=\frac{1}{\zeta^2-1}  \left[ \frac{m^3}{12 \lambda} \zeta (19-24 \zeta^2+9 \zeta^4) - c_2 (10 \zeta - 6 \zeta^3) \right] + c_1 (\zeta^2-1)+
3 c_2 (1-\zeta^2) \ln \frac{1+\zeta}{1-\zeta}
$$
with arbitrary constants $c_1$ and $c_2$. We fix $c_2=\frac{m^3}{12 \lambda}$
in order for our solution not to have singularities at $\zeta=\pm1$ (i.e.
$x=\pm \infty$).  So, the result for $f(x)$ is
$$ f(x) = - \cosh^{-2} \frac{m x}{2} \left(  \frac{3 m^3}{4 \lambda} \tanh \frac{m x}{2} +c_1 +\frac{m^4}{4 \lambda} x  \right)
$$
The solution $\phi(x)$ satisfies our kink condition  (\ref{kink}) for any $c_1$. It is easy to check that it 
does not change the classical kink energy (of course, this is true up to
$\frac{1}{M^4} $ order only). This indicates that $c_1$ is a zero mode for
our kink. It is really the first order in $\frac{1}{M^2}$ contribution
to the translational zero mode. 

In the rest of this paper we will be computing a lot of vacuum expectation 
values. This will be done using the propagators in trivial sector which read
\cite{shifman}
$$
\frac{M^2}{p_x^2+M^2} \frac{1}{p^2- W^{\prime \prime}(\phi_0)^2 \left(
1+ \frac{p_x^2}{M^2} \right)^{-2}}
$$
for bosons and 
$$
\frac{M^2}{p_x^2+M^2} \frac{\not{p}+ W^{ \prime \prime}(\phi_0) \left(
1+ \frac{p_x^2}{M^2} \right)^{-1} }{p^2- W^{\prime \prime}(\phi_0)^2 \left(
1+ \frac{p_x^2}{M^2} \right)^{-2}}
$$
for fermions. Actually, the only thing which is important to us is 
the ultraviolet behavior of these propagators. As in the momentum 
cut-off scheme, the great difference
with the computation in trivial sector will come because our
classical bosonic background is now an odd function of coordinate. 

\subsection{Supercurrent}

To obtain the supersymmetry current we extend Noether method for the case of higher-derivatives.
First, we integrate our Lagrangian by parts to rewrite it in the form with at most 
two derivatives on each field.

Consider some general Lagrangian $L(\phi,\partial_\mu \phi, \partial_\mu \partial_\nu \phi)$.
The Lagrange equation has the form
$$
 \frac{\delta L}{\delta \phi} - \partial_\mu 
 \frac{\delta L}{\delta (\partial_\mu \phi)}+
\partial_\mu \partial_\nu   \frac{\delta L}{\delta ( \partial_\mu \partial_\nu \phi)} = 0
$$
For local parameter $\epsilon=\epsilon(t,z)$ we write the 
variation of the $L(\phi,\partial_\mu \phi, \partial_\mu \partial_\nu \phi)$ 
as
\begin{equation}
\label{varl}
\delta L = \partial_\mu k^\mu + (\partial_\mu \bar{\epsilon} ) j^\mu
\end{equation}
where 
$k^\mu=\bar{\epsilon} k_0^\mu + (\partial_\nu \bar{\epsilon}) k_1^{\nu \mu} $
One may check directly that the current, defined this way,  is conserved on shell.
To do this, we note that 
\begin{equation}
\label{ll1}
 \delta L = \frac{\delta L}{\delta \phi} \delta \phi +  \frac{\delta L}{\delta (\partial_\mu \phi)} 
\delta (\partial_\mu \phi)+ \frac{\delta L}{\delta (\partial_\mu \partial_\nu \phi)} 
\delta (\partial_\mu \partial_\nu \phi)
\end{equation}
where we write $\delta \phi = \bar{\epsilon}  \Delta \phi $. Now, plug $k^\mu$ into (\ref{varl})
and equate terms with equal number of derivatives on $ \bar{\epsilon}$ in (\ref{varl}) and (\ref{ll1}).
The result is
$$
\partial_\mu k_0^\mu =   \frac{\delta L}{\delta \phi} \Delta \phi + 
\frac{\delta L}{\delta ( \partial_\mu \phi)} (\partial_\mu \Delta \phi)+
\frac{\delta L}{\delta ( \partial_\mu \partial_\nu \phi)} (\partial_\mu \partial_\nu \Delta \phi) 
$$
$$
k_0^\mu +j^\mu+\partial_\nu k_1^{\mu \nu} = 
\frac{\delta L}{\delta ( \partial_\mu \phi)}  \Delta \phi + 2
\frac{\delta L}{\delta ( \partial_\mu \partial_\nu \phi)}  \partial_\mu \Delta \phi
$$
$$
k_1^{\nu \mu}= \frac{\delta L}{\delta ( \partial_\mu \partial_\nu \phi)}  \Delta \phi
$$
Taking $\partial_\mu$ of second equation, and using the remaining two to express
derivatives of $k_0^\mu$ and $k_1^{\nu \mu}$, we obtain on shell $\partial_\mu j^\mu=0$,
the current is conserved.

The current is not unique though. Using this method one can compute, for example,
any of the following currents:
\begin{equation}
j_{PvN}^\mu = (\partial_\nu \phi) \gamma^\nu \gamma^\mu  \left( 1 - \frac{\partial^2_x}{M^2} \right) \psi + i W^\prime \gamma^\mu \psi 
+
\frac{\delta^\mu_x}{M^2} \left[ ( \Box \phi ) 
\stackrel{\leftrightarrow}{\partial_x}  \psi + 
i F {\stackrel{\leftrightarrow}{\partial_x}  } (\not \partial \psi) \right]
\ee
\be
j_{SVV}^{\mu} = j_{PvN}^\mu - \frac{1}{M^2} \epsilon^{\mu \nu } \partial_\nu
 (\not \partial \phi  \stackrel{\leftrightarrow}{\partial_x} \gamma^0 \psi)
\ee
\be
j_{OK}^{\mu}=j_{PvN}^{\mu} +\frac{1}{M^2} \epsilon^{\mu \gamma}
 \partial_\gamma 
\left[ (\partial_\nu\phi) \gamma^\nu \gamma^0 (\partial_x \psi) \right]
\ee
Clearly, the conservation of any of these current implies the
conservation of the others. The confusing issue for the last year
was in that if we use methods of \cite{shifman} for each of these currents
we compute different result for the anomaly. Below we will solve this
problem by analizing the contributions from the bosonic commutators.
At the same time, we will describe that only one of these three currents, 
$j_{OK}^\mu$ is regulated properly for loop computations. Nevertheless,
we will formally
compute the anomaly values from all three currents and observe that 
these values coincide.

For the  zero components of these currents we write
\footnote{Actually, it is easy to guess the $j_{OK}^{0}$ without doing
any computations: by the integration by parts the 
Lagrangian can be put in the form of sum of 
nonregulated lagrangian, which gives first two terms in $j_{OK}^{0}$,
and kinetic part of Lagrangian for fields $\partial_x \phi$, $\partial_x \psi$
and $ \partial_x F$. The susy transformations for that field differ
from susy transformations of prime fields by the term $\partial_x \epsilon$,
which would contribute only to the first component of the supercurrent,
so one may write the ``free'' supercurrent for this derivative fields in
$j_{OK}^{0}$ ,
this gives the third term.} 
\begin{equation}
j_{PvN}^0 = ( \partial_\nu \phi) \gamma^\nu \gamma^0  \left( 1 - \frac{\partial^2_z}{M^2} \right) \psi + i W^\prime \gamma^0 \psi
\end{equation}
\begin{equation}
j_{SVV}^{0} = \left( \partial_\nu   \left( 1 - \frac{\partial^2_z}{M^2} 
\right) 
\phi \right) \gamma^\nu \gamma^0  \psi + i W^\prime \gamma^0 \psi
\end{equation}
\be
j_{OK}^0 = ( \partial_\nu \phi) \gamma^\nu \gamma^0  \psi + i W^\prime \gamma^0 \psi + \frac{1}{M^2} ( \partial_\nu \partial_x \phi) \gamma^\nu \gamma^0  \partial_x \psi
\ee

\subsection{Supercharge Algebra and Central Charge VEV}

In this section we will compute central charge part of the supersymmetry algebra.
We define topological current $\zeta$ via
$$
\{j^\mu_\alpha, \bar{Q^\beta} \} = 2 (\gamma_\nu)_\alpha^\beta T^{\mu \nu} + 2 i (\gamma^5)_\alpha^\beta \zeta^\mu
$$
and the central charge is 
$$
Z = \int dz \zeta^0
$$
so that supercharges anticommute as (\ref{alg}). Note, that
expression (\ref{alg}) does not 
contain $\delta_\alpha^\beta$ term due to symmetry.
The corresponding term could still be present in current charge anticommutator
if it's zero component space integral is zero.

There are actually two equivalent ways to compute central charge. One is to use
the Poisson brackets directly
\be 
\left[ \phi(t,x),  \left( 1 - \frac{\partial^2_{x^\prime}}{M^2} 
\right) \dot{\phi}(t,x^\prime )
\right] = i \delta (x-x^\prime) 
\ee
\begin{equation}
\label{psic}
 \left\{ \psi_\alpha(t,x),  \left( 1 - \frac{\partial^2_{x^\prime}}{M^2} \right)  \psi_{ \beta} (t,x^\prime)
\right\} = \delta_{\alpha \beta} \delta(x-x^\prime). 
\label{fec}
\end{equation}
Another way is to note that $\delta j^\mu = [\bar{Q} \epsilon, j^\mu] $ and to use the
supersymmetry transformations for fields. Before doing so, one should make sure
that the charge $Q$ does really generate proper transformations of $\psi$ and $\phi$, which again requires the use of commutation relations.

Let us look at equation (\ref{fec}). The right hand side is an even 
function of $x-x^\prime$. Clearly, the left hand side should also
be symmetric with respect of exchanging of $x$ and $x^\prime$.
From this we conclude that we can change $\partial_x$
by $\partial_{x^\prime}$ in our operator. Is it very unusual for
quantum commutation relations to have nonlocal behavior? We remind
that exactly the same phenomenon occur in the sharp momentum cut-off
method of previous section. Possibly this is a general property of any 
regularization scheme which deserves future study. 

We are going to work out the bracket for all three currents,
but we want to start with $j_{SVV}^\mu$ because it was addressed by original
paper \cite{shifman}.
The central charge term in the anticommutator for $j_{SVV}^{0}$ has again
two parts, corresponding to cases 1) and 2) of the momentum cut-off
regularization computation, presented in the first chapter. The 
case 1) part comes from anticommutator of the fermionic fields multiplied
with bosonic loop, and has the form
$$
2 Z_{SVV,bos} = \int dx \int d x^\prime   \left( \left( 1 - \frac{\partial^2_x}{M^2} \right)  \partial_x \phi(x)
\right) W^\prime(\phi(x^\prime)) 
\left( 1 - \frac{\partial^2_{x^\prime}}{M^2} \right)^{-1} \delta(x-x^\prime) + 
$$
\be
\int dx \int d x^\prime   \left( \left( 1 - \frac{\partial^2_{x^\prime}}{M^2} \right)  \partial_{x^\prime} \phi(x^\prime)
\right) W^\prime(\phi(x)) \left( 1 - \frac{\partial^2_{x^\prime}}{M^2} \right)^{-1} \delta(x-x^\prime) 
\label{i1}
\ee
At this point we would like to give an example. Consider for a second
$ \int d y \left( f(y) \partial_y \delta(x-y) \right) $
There are two formal ways to compute this integral. One is integration by parts
which gives $f(y) \delta(x-y) |_{boundary} - \partial_x f(x)$. Another way
is to put $\partial_y \delta (x-y) = - \partial_x \delta (x-y) $ and 
switch the order of the integral and the derivative. If $x$ is
away from an integration interval, both give the same answer. If not,
the second way is wrong because $\int dy f(y) \delta (x-y) $ is not continuous
function of $x$, it has a jump when $x$ crosses the boundary of the 
integration 
region. If we would use the second method for (\ref{i1}), we would get \cite{shifman}
$$
Z_{SVV,bos}=\frac{1}{2} \left[ W^\prime \partial_x \phi + \left( \left( 1 - \frac{\partial^2_x}{M^2} \right) \partial_x \phi \right) 
\left( \left( 1 - \frac{\partial^2_x}{M^2} \right)^{-1} W^\prime \right) \right] 
$$
We consider this intermediate result of \cite{shifman} as being wrong. 

On the other hand, just notice that
we can change $x^\prime$ by $x$ in the last $\left( 1 - \frac{\partial^2_{x^\prime}}{M^2} \right)^{-1} $, and then both terms become equal. I.e.,
$$
2 Z_{SVV,bos} = 2 \int dx \int d x^\prime   \left( \left( 1 - \frac{\partial^2_x}{M^2} \right)  \partial_x \phi(x)
\right) W^\prime(\phi(x^\prime)) 
\left( 1 - \frac{\partial^2_{x^\prime}}{M^2} \right)^{-1} \delta(x-x^\prime)  
$$
(Note, that if we would apply the ``second method'' to this result we would 
get the classical answer, without any anomaly at all!).

Here we need to define what we mean by  $\left( 1 - 
\frac{\partial^2_{x^\prime}}{M^2} \right)^{-1} \delta(x-x^\prime) $
 and by $  \delta(x-x^\prime) $ itself. If we had imposed boundary conditions
for the  quantum fields, we would need to redefine Dirac delta function as a sum of the modes with corresponding boundary conditions, as described in \cite{misha}. 
Nevertheless, we try to avoid imposing boundary conditions here, following the
ideas of \cite{shifman}. So, we can just use $ \delta(x-x^\prime) = \int \frac{dq}{2 \pi}
\exp(i q (x-x^\prime))$ and then the regularization procedure spreads it to
\be
\hat{\delta}(x-x^\prime)=\left( 1 - 
\frac{\partial^2_{x^\prime}}{M^2} \right)^{-1} \delta(x-x^\prime) = \int_{- \infty}^{+ \infty} \frac{dq}{2 \pi} 
\frac{\exp(i q (x-x^\prime))}{1+\frac{q^2}{M^2}} = \frac{M}{2} 
\exp(-M |x-x^\prime|)
\ee
Now we can rewrite the VEV of $Z_{SVV,bos}$ in the form, analogous 
to (\ref{metka1}), and the anomaly is given by the term 
\be \int_{-L}^{L} dx \int_{-L}^{L} dx^\prime 
W^{\prime \prime}(x)
\langle \eta(x) \partial_{x^\prime} \left( 1 - 
\frac{\partial^2_{x^\prime}}{M^2} \right) \eta(x^\prime) \rangle \hat{\delta}(x-x^\prime)
\label{metka2}
\ee
Actually, the effect of regularization on the propagator is removed
by the factor $\left( 1 - 
\frac{\partial^2_{x^\prime}}{M^2} \right) $ in the correlator. Effectively,
we get nonregularized propagator. The loop integral is then divergent
at all and there is no fare way to compute such a thing. In the same
time, one is tempted to cheat: let us think that the loop integral runs 
up to some finite value of momentum, and then we may first take $x$ integrals,
and after that put that momentum to infinity. If we do that, it shouldn't
surprise the reader that we get the result which we argued to be 
correct in section 1. 
(In this case the fermionic contribution is zero.)
We just have the old
situation with under-regulated propagator, with exactly the same 
consequences. Thus, though the current $j^\mu_{SVV}$ is not regulated properly
by our scheme and that is why it  should not be used for further computations, the result
that it gives is formally correct. 

Now we move to the current $j_{PvN}^\mu$. The computation with this current is 
easy, and the resulting integral for an anomaly is
\be \int_{-L}^{L} dx \int_{-L}^{L} dx^\prime 
W^{\prime \prime}(x)
\langle \eta(x) \partial_{x^\prime}  \eta(x^\prime) \rangle {\delta}(x-x^\prime)
\label{metka3}
\ee
which gives well regulated propagator, but, unfortunately, nonregularized
delta function. Here, of course, the result for an anomaly coming from
the bosonic sector is zero ( as it
has to be in the case where delta function is under-regulated). But
the fermionic sector contributes now and the total anomaly is again the
correct one. Thus, though
the current $j_{PvN}^\mu$ does not supply well regulated expression either,
the formal result for the anomaly is the same.

Now we move to the winning one, the current $j_{OK}^\mu$.
The first two terms in it give the well regulated expression
\be \int_{-L}^{L} dx \int_{-L}^{L} dx^\prime 
W^{\prime \prime}(x)
\langle \eta(x) \partial_{x^\prime} \eta(x^\prime) \rangle \hat{\delta}(x-x^\prime)
\label{metka4}
\ee
Using our trick for the substitution of kink shape by a step shape  
(see Appendix A ) , we arrive at the following expression for this integral
\be
W^{\prime \prime}(L) \frac{iM}{2} \int_{-L}^{L} dx \hspace{0.2cm} {\rm sign}(x) \int_{-L}^{L} dx^{\prime}
\int \frac{d^2p}{(2\pi)^2} \frac{- i p_x}{\left(1+\frac{p_x^2}{M^2} \right) p^2}
\exp(-M|x-x^\prime| +ip(x-x^\prime))
\label{int2}
\ee
which is easy to compute (see Appendix B ) the result for $L \to \infty $ 
being 
\be
\label{rezz}
\frac{W^{\prime \prime}(\infty)}{4 \pi}
\ee
which is the half of the anomaly. Clearly, another half is coming from the 
fermionic loop, which correspond to case 2) of the first chapter. The
resulting integral is exactly equal to (\ref{metka4}), so that the total
anomaly is again  $$ \frac{W^{\prime \prime}(\infty)}{2 \pi}$$

It remains to analyze the last term in the $j_{OK}^\mu$. This term
gives the following contribution to the central charge anomaly
$$
W^{\prime \prime}(L) \frac{1}{M^2} \int_{-L}^{L} dx \hspace{0.2cm} {\rm sign}(x) \int_{-L}^{L} dx^\prime
\int \frac{d^2p}{(2\pi)^2} \frac{p_x^2 \exp(ip_x(x-x^\prime))}{p^2 \left( 
1+\frac{p_x^2}{M^2} \right)}  \int \frac{dq}{2 \pi} 
\frac{q \exp(iq(x-x^\prime)}{\left( 
1+\frac{p_x^2}{M^2} \right) }
$$ 
An easy computation shows that this integral is zero (for $x \ne x^\prime$,
both momentum integrals converge and we can change the order of the 
integration, the result is zero. For $x=x^\prime$, change in the order of 
the integration is not needed as the last momentum integral is seen to be zero
on its own).

This concludes our treatment of an anomaly in the higher-derivative 
regularization. 

We would like also to comment on the supercharges which follow from the
currents which we have discussed.  If one writes the
supercharges in components, following \cite{witten}, 
then the only one supercharge, 
that looks 
as though it would have zero second component, 
is $Q_{SVV}$, namely
$$
Q_{SVV,1}=\int dz \left[ 
\left( \left( 1 - \frac{\partial^2_z}{M^2} \right) \partial_t \phi \right) \psi_1+ 
\left( \left( 1 - \frac{\partial^2_z}{M^2} \right) \partial_z \phi \right) \psi_2+W^\prime \psi_2
\right]
$$
$$
Q_{SVV,2}=\int dz \left[ 
\left( \left( 1 - \frac{\partial^2_z}{M^2} \right) \partial_t \phi \right) \psi_2+ 
\left( \left( 1 - \frac{\partial^2_z}{M^2} \right) \partial_z \phi \right) \psi_1-W^\prime \psi_1
\right]
$$
and  $\psi_1$ term in $Q_{SVV,2}$ reproduces the Bogomol'nyi 
equation (\ref{kink}). Nevertheless, the quantum loops with this
current are not regulated, as we have described above. 

Finally, we would like to comment on another approach to the anomaly computation based 
on the dimensional regularization
scheme (DREG) 
and the one given in \cite{shifman}.
We are confident that if we could repeat our momentum cut-off 
program here, we would then get the same value for the anomaly. Nevertheless,
after reading our paper and the paper \cite{shifman} one may be slightly confused:
is the anomaly from momentum cut-off and higher-derivative scheme, presented 
above, the same as the dimensional regularization anomaly of \cite{shifman}? Our answer
is ``no''. We feel that the anomaly in DREG should arise exactly the same
way as it does in other schemes, from the same integral. 

The computation of \cite{shifman} says that
$$
(\gamma^\mu J_\mu) = (D-2) (\partial_\mu \phi) \gamma^\mu \psi - Di W^\prime \psi
$$
and using
$$
j^0=\partial_\mu \gamma^\mu \gamma^0 \psi +i W^\prime \gamma^0 \psi
$$
one easily computes
$$\left\{ (\gamma^\mu J_\mu)_{anom} (z), \bar{j}^0(z^\prime) \right\}=
(D-2) \partial_\mu\phi(z) \gamma^\mu \gamma^\nu (\partial_\nu \phi)
\hat{\delta}(z-z^\prime) + $$ $$i W^\prime (D-2) (\partial_\mu \phi) 
\gamma^\mu 
\hat{\delta}(z-z^\prime) - i D W^\prime(\partial_\nu \phi) \gamma^\nu \hat{\delta}(z-z^\prime)
$$
it is only at this point when one must take quantum
average, and we observe that $D-2$ with $-D$ give just $2$. No anomaly
is seen on this level! 

The detailed treatment of the dimensional regularization scheme was
given by \cite{vse}. The recent paper \cite{noah} has proven that the
counterterm, used by \cite{vse} is really equal to the tadpole graph
contribution, and thus removes tadpoles completely. This is the specific
property of dimensional regularization. If one thinks of the results 
of \cite{vse} in the context of another regularization scheme, the 
counterterm needed may be different. For example, if one thinks
of the momentum cut-off scheme, the traditional counterterm
of \cite{peter} and \cite{misha} differs from the one used in \cite{vse}
by the integration by parts, i.e. by the finite boundary term. Nevertheless,
this two counteterms seem to coincide when dimensionally regulated.
We find this miraculous.

 We stress that the results of \cite{vse} are
only the dimensional regularization results.  
It is well known, that the naive momentum cut-off gives wrong results
for mass, the reason for that we have described in \cite{we}. On the same time,
it is the momentum cut-off that is being implicitly assumed in the
mass computation of \cite{vse} when the authors unite different
sums/integrals over the continuous spectrum under one integral sign replacing
the densities of states by derivative of phase shift. 
It is easy to check that the usage of the wrong counterterm brings
in exactly that finite correction to the mass, that was lost when doing 
the momentum cut-off. So the final result looks to be correct. This
may be not the coincidence and there still remains something to be 
studied.   

We are also surprised by the way which \cite{vse} uses to compute 
the one loop correction for the central charge. This method does
not disclose the anomalous nature of the correction, and it sounds
to us as the computation of $\langle Z \rangle = \langle H \rangle +
\langle Q_2^2 \rangle $, i.e. we feel that what is really proved is 
that $ \langle Q_2^2 \rangle $ is zero. This and other issues about
the phase-shift method will
be discussed elsewhere \cite{we}. 

\section{Conclusions}

In this paper we have developed a new approach to the momentum 
cut-off regularization,
which we call {\it generalized} momentum cut-off regularization scheme. 
This new technique resolves the 
issue of the one loop anomaly in the central charge of the 
supersymmetric kink with
a sharp momentum cut-off regularization. We have imposed a cut-off 
not only for the 
loop integrals (this cut-off we call $\Lambda$), but also for the Dirac delta functions themselves (the cut-off that we call $K$). 
The consistency required that these
two cut-offs are equal. We emphasize that if they were not, such a theory would be internally inconsistent. 
In particular, if $\Lambda>K$ then such a theory contains ``purely classical'' modes in operator expressions, i.e. the modes multiplied 
by perfectly commuting (anticommuting) creation/absorption ``operators'', and this gives a different result
for an anomaly. On the other hand, if $\Lambda<K$ then the commutators for quantum fields are inconsistent, because
then for fields $\phi(x) \sim \exp(iqx)$ $K>q>\Lambda$ the right hand side of commutation relations can be nonzero and the 
left hand side can not.

Note that imposing momentum cut-off on the Dirac delta functions makes
them nonlocal. The same effect happens in the higher-derivative regularization scheme,
and may be possibly a common feature of regularized quantum field theories. It is not
immediately clear to us that this nonlocality follows from the nonlocality
of quantum propagators, which is, of course, the meaning of regularization.
The direct consequence of this is that the (anti)commutators of different symmetry currents 
result in currents that are not local but spread over the finite interval of the size $\sim 1/\Lambda$.
In particular, the topological current that results from the susy algebra is not local. Again, this
is quite natural once we start from the very beginning with the model where there are no local
objects -- of course, one can not think about perfectly localized quantum field if only the
Fourier harmonics up to $\Lambda$ are allowed. 

We have also elaborated on the higher-derivative regularization scheme 
of \cite{shifman}, 
as used for the supesymmetric kink. We gave 
consistent technical prescriptions for
this method and we have computed the correct value of an anomaly in this scheme. 
The reason that we dwell on this topic is that the scheme is preserving supersymmetry
manifestly, and we find it useful for further investigations of the higher-dimensional
problems. 

In the higher-derivative regularization scheme we notice that the susy current
is not unique -- one may change it by adding a full derivative term. Not all the
currents produced this way are regulated by the scheme, so one should pick only a 
very special current, namely one  for which one computes from the susy 
algebra a regularized topological current, i.e. the topological current
such that when one takes the vacuum expectation value of it,
the regularization effect in propagator  
is not removed by extra powers of momentum in the vertex.
The topological current that we obtain this way is not local. The nonlocality is the reason that the anomaly is present. 
We conclude that it is essential that the topological currents are not local in any regularization scheme, otherwise
no anomaly would be seen. It would be definitely very instructive to see how does this nonlocality work in the
case of the dimensional regularization, but in this paper we only comment on the earlier approaches to the
dimensional regularization and argue that in our opinion the problem is still not solved.

We have checked that the properly regulated supersymmetric kink system has the central charge anomaly which 
is exactly equal to the one loop mass correction computed by
\cite{misha},\cite{vse},\cite{dune}. This concludes the discussion on BPS bound saturation. We
have worked out in great detail two regularization schemes, but the conclusion looks
to be quite general: There are only three results possible. If one
gets the zero or twice the proper anomaly results, this means that the
theory does not regulate properly the loops for the superalgebra. The anomaly
always comes from the nonlocal topological currents due to the integration
in the vicinity of the boundary. I.e. in the properly regulated theory 
the boundary itself is smeared over 
the finite interval. The length of this interval is defined by the inverse 
of the regulator. Nothing can be left local in the theory which is regulated properly, neither
the position of the boundary, nor the delta functions in the canonical commutation relations.  
\\ \\

{\bf Acknowledgments:} Through all my work, I have always felt greatest support of
my parents, Vitalii and Natalia. This work would be impossible without ideas, 
help and encouraging by my teacher and advisor,
Peter van Nieuwenhuizen. I benefited a lot from the collaboration
with Alfred Goldhaber and Misha Stephanov, and from the discussions with Bogdan Kulik. 
I am very grateful to 
M.Shifman, A.Vainshtein and M.Voloshin for very useful 
critical discussion of some of the results presented here.
It is a pleasure to 
acknowledge the discussions and correspondence with E.Farhi, N.Graham, 
R.L.Jaffe, A.Rebhan, C.Vafa 
and H.Weigel. I am very grateful to Liliya Vugmeister for proof-reading the
draft of this paper.

\appendix
\section{Computation of the integral for the anomalies in the momentum cut-off regularization}  
   
The computation of the integral (\ref{int1}) is easy. Notice, that 
if we would replace the kink background with the
step like shape: $W^{\prime \prime}(x) \to 
W^{\prime \prime}(L) {\rm sign}(x)$, 
then we would compute the same answer for this integral. There is good
reason for that: if we put $L=\infty$ from the very start, we see that
this integral is just zero. I.e., the integral in the limits $(-L,L)$ is
equal to the negative of the same integral over the region $|x|>L$, and for
sufficiently large $L$ such substitution is justified. We will be using
the same trick when working with the higher-derivative regularization 
integrals. 

We start with the $p_0$ integral in (\ref{int1}). Remember
that we put momentum cut-off on $p_x$ integral, while we integrate in
$p_0$ over all real axes. The integral can be performed with residues 
or by Wick rotation, either way gives 
\be
\frac{i}{8\pi^3} 2 \pi i \frac{i}{2} W^{\prime \prime}(L)  
\int_{-L}^L dx \int_{-L}^L dx^\prime \int_{-\Lambda}^{\Lambda} dp
\int_{-K}^K {\rm sign} (x) {\rm sign} (p) \exp(i (x-x^\prime) (p+q)
\ee
(we drop subindex $x$ of $p_x$). Next, we take the coordinate 
integrals to obtain
\be
-\frac{i}{8 \pi^2}  W^{\prime \prime}(L)  \int_{-\Lambda..\Lambda} dp
\int_{-K}^K dq  2 i  {\rm sign} (p) \frac{2 \sin (L(p+q)) - \sin(2L(p+q)) }{(p+q)^2}
\ee
Integral over $p$ from $-\Lambda$ to $\Lambda$ is clearly twice the integral 
from $0$ to $\Lambda$. We introduce $z=p+q$ and get
\be
-\frac{i}{8 \pi^2}  W^{\prime \prime}(L) 2 \int_{0}^{\Lambda} dp
\int_{p-K}^{p+K} dz  2 i   \frac{2 \sin (Lz) - \sin(2Lz)) }{z^2}
\ee
next, we do $z$ integral which yields
\be
 \frac{1}{2 \pi^2}  W^{\prime \prime}(L) \int_0^\Lambda dp 
\label{uh101}
\ee
 $$ \left( \frac{
(- 4 \sin(Lp) \cos(LK) +2 \sin(2Lp) \cos(2LK) )K }{K^2-p^2}
\right. $$
$$ \left.
+ \frac{
(4 \cos(L p) \sin(LK) - 2 \cos(2Lp) \sin(2LK) )p  
}{K^2-p^2} \right. $$ $$ \left.
+ L ({\rm Ci }((p+K)L) - {\rm Ci }(2(p+K)L)-  {\rm Ci }((p-K)L)+
 {\rm Ci }(2(p-K)L) )
\right) 
$$
where the function ${\rm Ci}(x) = \gamma +\ln (x) +\int_0^x 
\frac{\cos t -1}{t} dt $ is integral cosine function. If $K>\Lambda$ then
no terms in this expression are singular. One may compute the expressions,
but it is clear already now that they will come in pairs such that 
for any term with $L$ there is a counterpart with $2L$, so taking
$L \to \infty$ limit will cancel all the terms and the answer is zero. 
The situation is different for $K \le \Lambda$.
The last line in (\ref{uh101}) is still zero - it is clear
from the definition of ${\rm Ci(x)}$: 
The singularities in ${\rm Ci}$ are logarithmic ( using that $\int \ln x  dx = x \ln x  -x +const$, we see that each of them separately would give
finite contribution) and for the last line
of (\ref{uh101})
they are seen to cancel for any given $L$.  
At the same time, the
infrared singularity develops in the fractions, the first two terms of  
(\ref{uh101}). We can not argue about the value of the integral of these
fractions until we take it explicitly. This is easy to do, the result being
$$
 \frac{1}{2 \pi^2}  W^{\prime \prime}(L) 
\left( 
{\rm Si}(2 \Lambda L +2 KL) - 2 {\rm Si} (LK-\Lambda L) - \right. $$ 
$$ \left.
 2 {\rm Si} (\Lambda L +K L) + {\rm Si} (-2 \Lambda L+2KL) - 2 {\rm Si}(2KL)
+4 {\rm Si} (KL)
\right) 
$$
with ${\rm Si}= \int_0^x \frac{\sin t}{t} $ being the integral sine. The 
limit $L=\infty$ is then trivial and our final result is
\be
\frac{W^{\prime \prime}(\infty)}{4 \pi} \left( 1+{\rm sign} (\Lambda-K ) \right)
\ee
which is exactly the result (\ref{polotveta}).

\section{Computation of the integral for the anomalies in the higher-derivatives regularization}  
Here we compute (\ref{int2}). This computation is much easier than one in the momentum 
cut-off scheme.
First, take $p_0$ integral, then $x$ and $x^\prime$, and finally $p_x$  integrals to get 
$$
-\frac{W^{\prime \prime}(L)}{2 \pi} \left( 
\exp(3ML) (8-4ML) -4 \exp(4ML) + \right. $$ 
$${\rm Ei}(1,ML) \exp(4ML) (2-4ML+4(ML)^2) +{\rm Ei}(1,2ML) \exp(4ML) (-1+4ML-8(ML)^2)+ 
$$
$$
\left.
4\exp(2ML)(ML-1)+{\rm Ei}(1,-2ML) - 2 {\rm Ei }(1, -ML) \exp(2ML)
\right) \exp(-4ML)
$$
where ${\rm Ei} (n,x)=\int_1^\infty \frac{\exp(-xt)}{t^n} dt$ is integral exponent function. 
Taking the limit $L \to \infty$ gives the result (\ref{rezz}). 

 Notice that to compute this result we have substituted the exact kink shape by the sign function. As we have described already, technically the reason is that the
nonzero value of this integral comes from the regions around the boundary. This
means that the boundary is not localized at one point. It is smeared over the
interval of the length $1/M$. Nothing local can be left in the properly 
regulated theory!

\end{document}